\documentclass[12pt]{article}
\usepackage{graphicx}
\begin{document}

\title{Global polarization and parity violation study in Au+Au collisions}
\author
{
Ilya Selyuzhenkov for the STAR Collaboration\\
{\it \small Department of Physics and Astronomy, Wayne State University, USA} \\ 
{\small {\rm seluzhen@physics.wayne.edu}}
}
\date{}
\maketitle
\begin{abstract}
We present results on the parity violation effects and global system polarization measurements
in Au+Au collisions at $\sqrt{s_{NN}}$ = $62$ GeV obtained with the STAR detector at RHIC.
The parity violation effects are studied by three particle azimuthal correlations of charged particles.
The global polarization of the system is examined by measuring the polarization of strange hyperons 
with respect to the collision reaction plane.


\end{abstract}

\section{Introduction}

The system created in non central relativistic nucleus-nucleus collision possesses large angular orbital momentum.
Some of the most interesting and important phenomena predicted to occur in such a system are
the strong parity violation \cite{Kharzeev2004} and the global system polarization \cite{Liang}.
Strong $P$ and $CP$ violation should be revealed via preferential emission
of the same charge particles in the direction along the system angular momentum.
This effect is already known for a few years \cite{Kharzeev1998}-\cite{Finch2002}
but only recently clear theoretical estimates and experimental observables were suggested \cite{Voloshin}.
The effect of the global system polarization was first discussed in \cite{Liang}
where theoretical arguments for the globally polarized system
produced in non central nucleus-nucleus collision were presented
and possible observable consequences and theoretical estimates were given.
The global polarization originates from transformation of the orbital angular momentum of the system into the particle's spin.
The latter leads to the polarization of secondary produced particles along the system orbital momentum.

The effects of the parity violation and the global system polarization
are both defined by the presence of the large angular momentum in the colliding system.
Since the angular momentum of the system is perpendicular to the collision reaction plane
the effects under consideration can be studied by methods and techniques developed for anisotropic flow analysis.
In this paper the parity violation effects are studied by three particle azimuthal correlations of charged particles.
This technique is based on the mixed harmonic method  \cite{Poskanzer1998,Borghini2002}.
The global polarization of the system is examined by measuring the polarization of strange hyperons
with respect to the collision reaction plane.
The hyperon polarization can be assessed via anisotropy in angular distribution 
of decay products in the hyperon's rest frame measured with respect to the collision reaction plane.
This measurement is similar to the standard method of extracting directed flow \cite{Poskanzer1998}.

\section{Global polarization of hyperons}

Particles produced in the system with large angular orbital momentum
are predicted to be polarized along the collision reaction plane direction due to spin-orbital coupling \cite{Liang}.
Such polarization can be defined from the angular distribution of hyperon decay
products with respect to the collision reaction plane:
\begin{eqnarray}
\nonumber
\frac{dN}{d \cos~\theta^*} \sim 1~+~\alpha~P_{H}~\cos~\theta^*~~,
\end{eqnarray}
where $P_H$ is the hyperon polarization with respect to the collision reaction plane,
$\alpha$ is the hyperon decay constant ( $\alpha = 0.642$ for $\Lambda \to p~+~\pi^- $), and
$\theta^*$ is the angle between the normal for the collision reaction plane and
the nucleon 3-momentum in the hyperon's rest frame.

Global polarization $P_{H}$ can be defined from two particle azimuthal correlations
with respect to the collision reaction plane:
\begin{eqnarray}
\nonumber
P_{H}~=~\frac{8}{\pi\alpha}\langle \sin \left( \varphi - \Psi_{RP}\right)\rangle~~,
\end{eqnarray}
where $\varphi$ is the azimuthal angle of hyperon's decay product 3-momentum in the hyperon's rest frame,
and $\Psi_{RP}$ is the collision reaction plane angle.

\begin{figure}[t]
\begin{center}
\includegraphics[width=0.5\textwidth]{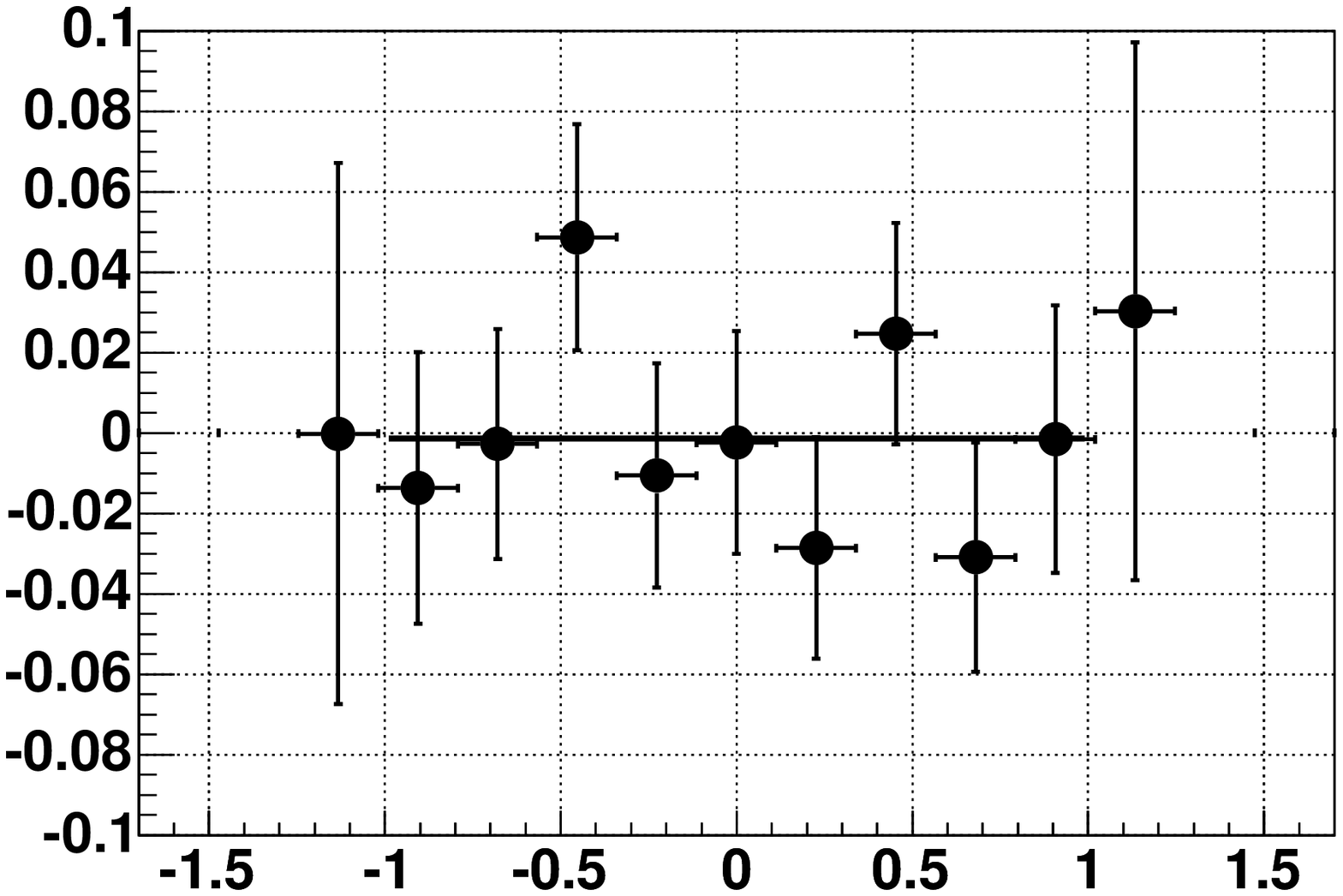}
\put(-250,70){\rotatebox{90}{$P_{\Lambda}$}}
\put(-170,122){{\bf STAR Preliminary}}
\put(-115,-5){$\eta_{\Lambda}$}
\caption{\label{lambdaPolarization}
$\Lambda$ polarization as a function of pseudo-rapidity for centrality 10-70\%.
Black points are STAR preliminary data for Au+Au at $\sqrt{s_{NN}} = 62$ GeV.
The line is a constant line fit to experimental data (see text for more details).
}
\end{center}
\end{figure}

Fig.\ref{lambdaPolarization} shows $\Lambda$ polarization as a function of $\Lambda$ pseudo-rapidity.
In this calculation the reaction plane is obtained from directed flow of
charged particles in the pseudo-rapidity range $2.7 < |\eta| < 3.9$ from STAR Forward Time Projection Chamber (FTPC)
and from fragmentation neutrons with $|\eta|>6.3$ from STAR Zero Degree Calorimeter-Shower Maximum Detector (ZDC-SMD).
Points in Fig.\ref{lambdaPolarization} represent the STAR preliminary results 
from the experimental data for Au+Au collisions at $\sqrt{s_{NN}}=62$ GeV.
Constant line fit to the data gives the upper limit for $\Lambda$ polarization: $P_{\Lambda} ~= ~(-1.44~\pm~9.66) \times 10^{-3}$.
The obtained results in the errors range are consistent with zero.
The theoretical prediction \cite{Liang} for the $\Lambda$ polarization is $P_{\Lambda} ~= ~- 0.3$
with the magnitude far larger than that observed in the current measurement.

\section{Parity violation in hot QCD}

Parity violation, discussed in \cite{Kharzeev2004}, implies asymmetry in the particle emission along the system orbital momentum.
Such asymmetry fluctuates event by event but it relative sign for the opposite charge particles stays constant.
This results in the following azimuthal distribution \cite{Voloshin}:
\begin{eqnarray}
\frac{dN_{\pm}}{d \varphi} \sim 1~+~2 a_{\pm}~\sin~\varphi~~,
\label{parityDistribution}
\end{eqnarray}
where $\varphi$ is the particle emission azimuthal angle relative to the reaction plane, $a_{\pm}=(4/\pi)~Q/N_{\pm}$ is the 
asymmetry\footnote{$a_{\pm}$ is related to the asymmetry $A_{\pm}$ defined in \cite{Kharzeev2004} via: $a_{\pm}= (4/\pi)~A_{\pm}$} 
in charged particle production,
defined by the ratio of topological charge $Q$ ($|Q|\geq 1$) to the corresponding charged particle multiplicity $N_{\pm}$.

It was shown in \cite{Voloshin} that one can measure charged particle anisotropy (\ref{parityDistribution})
by using mixed harmonic method \cite{Poskanzer1998,Borghini2002} applied in a symmetric pseudo-rapidity region:
\begin{eqnarray}
\nonumber
a_i a_j~=~- \langle \cos\left( \varphi_i +\varphi_j - 2 \Psi_{RP} \right)\rangle~~,
\end{eqnarray}
where $\varphi_{i,j}$ is the azimuthal angle of particles/antiparticles (for example $\pi^+$ and $\pi^-$), 
and $\Psi_{RP}$ is the collision reaction plane angle.

\begin{figure}[t]
\begin{center}
\includegraphics[width=0.55\textwidth]{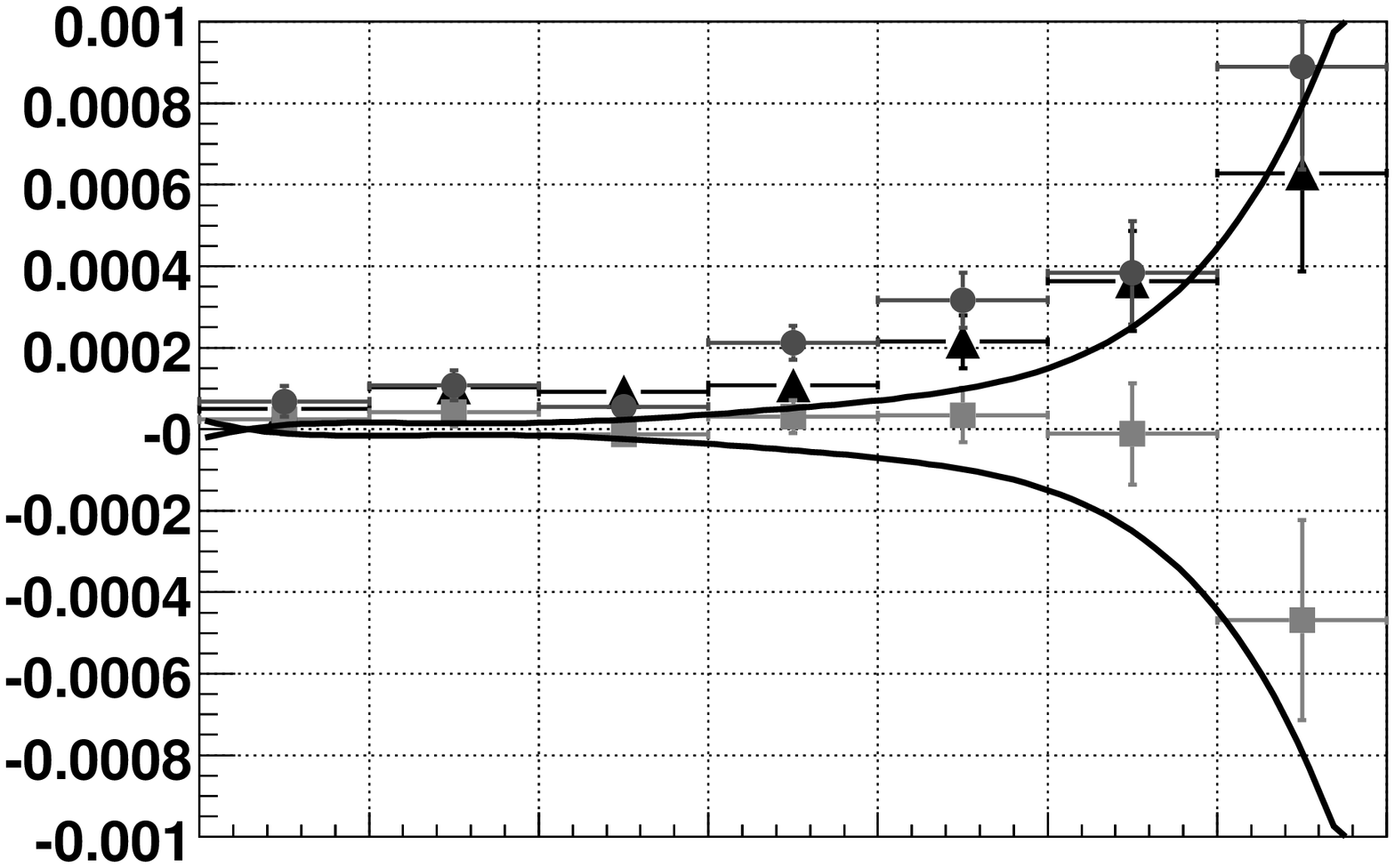}
\put(-178,125){{\bf STAR Preliminary}}
\put(-130,-8){$\sigma/\sigma_{tot}$, \%}
\put(-201, 7){\tiny\bf 0-5\%}
\put(-177, 7){\tiny\bf 5-10\%}
\put(-150, 7){\tiny\bf 10-20\%}
\put(-121, 7){\tiny\bf 20-30\%}
\put(-93, 7){\tiny\bf 30-40\%}
\put(-65, 7){\tiny\bf 40-50\%}
\put(-35, 7){\tiny\bf 50-60\%}
\put(-260,50){\rotatebox{90}{$a_{\pm}^2~,~~ a_{+}a_{-}$}}
\caption{\label{parityCheck}
Charged particle asymmetry parameters as a a function of standard STAR centrality bins
selected on the basis of charged particle multiplicity in $|\eta| < 0.5$ region.
Points are STAR preliminary data for Au+Au at $\sqrt{s_{NN}} = 62$ GeV:
circles are $a_{+}^2$, triangles are $a_{-}^2$ and squares are $a_{+}a_{-}$.
Black lines are theoretical prediction \cite{Kharzeev2004} corresponding to the topological charge $|Q|=1$.
}
\end{center}
\end{figure}

Fig.\ref{parityCheck} shows the charged particle asymmetry parameter (symmetric pseudo-rapidity range $|\eta| < 0.5$) for different centrality of the collision.
In this analysis the reaction plane is obtained from 
elliptic flow of charged particles with pseudo-rapidity $|\eta|<1.3$ from STAR Time Projection Chamber (TPC).
Points represents the STAR preliminary results from the experimental data for the Au+Au collisions at $\sqrt{s_{NN}}=62$ GeV.
The lines are theoretical predictions \cite{Kharzeev2004} for the minimum value of the topological charge $|Q|=1$.

\section{Summary}
The full statistics for the STAR Au+Au data at $\sqrt{s_{NN}} = 62$ GeV for strange hyperons have been analyzed.
The obtained upper limit for the global polarization of $\Lambda$ hyperon, $P_{\Lambda} ~= ~(-1.44~\pm~9.66) \times 10^{-3}$,
is far below the value predicted in \cite{Liang}.

Although the systematic uncertainties have not been yet clearly understood,
the current analysis is capable of setting an upper limit
on the effect of strong $C$ and $CP$ violation in nucleus-nucleus collisions discussed in \cite{Kharzeev2004}.

\end{document}